\documentclass[russian,english]{article}
\usepackage[russian,english]{babel} 
\usepackage{cite}

\def\rrr#1\\{\par
\medskip\hbox{\vbox{\parindent=2em\hsize=6.12in
\hangindent=4em\hangafter=1#1}}}

\begin{document}

\selectlanguage{english}

\begin{center}
{\Large{\bf On The Painlev\'e Property For A Class Of Quasilinear Partial Differential Equations}}\\*[3mm]

Stanislav Sobolevsky\\
New York University, sobolevsky@nyu.edu

\end{center}

The last decades saw growing interest across multiple disciplines in nonlinear phenomena described by partial differential equations (PDE). Integrability of such equations is tightly related with the Painlev\'e property - solutions being free from moveable critical singularities. The problem of Painlev\'e classification of ordinary and partial nonlinear differential equations lasting since the end of XIX century saw significant advances for the equation of lower (mainly up to fourth with rare exceptions) order, however not that much for the equations of higher orders. 

Recent works of the author have completed the Painlev\'e classification for several broad classes of ordinary differential equations of arbitrary order, advancing the methodology of the Panlev\'e analysis. This paper transfers one of those results on a broad class of nonlinear partial differential equations - quasilinear equations of an arbitrary order three or higher, algebraic in the dependent variable and including only the highest order derivatives of it.
Being a first advance in Painlev\'e classification of broad classes of arbitrary order nonlinear PDE's known to the author, this work highlights the potential in building classifications of that kind going beyond specific equations of a limited order, as mainly considered so far.

\section{Introduction}

Nonlinear Partial Differential Equations and especially evolutionary PDE's, describing dynamics of nonlinear phenomena saw increasing interest across different fields of physics, such as statistical mechanics, fiber optics, fluid dynamics, condensed matter, elementary particle physics, astrophysics as well as reaction-diffusion systems in chemistry and competition of species in biology \cite{musette1999painleve}.

Painlev\'e property plays important role in the analysis of PDE's as being tightly connected with their integrability. Equations with Painlev\'e property often happen to be integrable either analytically either through inverse scattering transform \cite{ablowitz1991solitons}. Moreover, famous Ablowitz-Ramani-Segur conjecture \cite{ablowitz1980connection} suggests that all reductions of PDE's integrable by means of inverse scattering transform should possess the Painlev\'e property. On the other hand, when one can not find the way to integrate equation having Painlev\'e property, it often warrants even higher interest. This is because solutions of such an equation could be seen as a source for new single-valued or meromorphic functions, such as second-order Painlev\'e transcendents (see for instance \cite{ince1956}, chapter XIV) and their higher-order generalizations \cite{kudryashov1997first}. Additional practical interest to the equations with Painlev\'e property is warranted by the fact that their solutions often demonstrate interesting properties, such as solutions of Korteweg-de-Vries \cite{korteweg1895xli} or Schr\"odinger \cite{schrodinger1926undulatory} equations being able to describe propagation of bell-shaped solitary waves \cite{zabusky1965interaction,  ablowitz1991solitons, musette1999painleve} or even collisions of multiple solitons \cite{hirota1971exact}.

The Painlev\'e classification of ordinary differential equations is one of the long-lasting problems of analytic theory of differential equations rooted in the end of XIX century \cite{ince1956}. For PDE's the Painlev\'e has been defined in the 1980's by Weiss, Tabor And Carnevale in \cite{weiss1983painleve}.

In spite of numerous achievements on the classification of the equations of some limited order (mainly up to fourth) over the last more than hundred years, the general problem for the higher order equations remains unsolved. This paper contributes towards solution of the arbitrary-order problem by proving absence of the Painlev\'e property classification for a broad class of quasilinear PDE's.

\section{Recent advances in Painlev\'e classification of ordinary and partial differential equations}

The Painlev\'e classification of the non-linear ordinary differential equations (ODE)
\begin{equation}
w^{(n)}=F(w^{(n-1)},w^{(n-2)},...,w,z),
\label{ODE}
\end{equation}
is mainly known for the order $n\leq 4$. For the order $n=1$ the necessary and sufficient condition of the Painlev\'e property in case of the right-hand side $F$ algebraic in $w$ is a well-known classical result (see for instance \cite{ince1956}, chapter XIII). For the order $n=2$ classification of the equations with the rational right-hand side $F$ has been built in the classical works of Painlev\'e and Gambier (see for instance \cite{ince1956}, chapter XIV). For the order $n=3$ classification has been
started in the famous work of Chazy \cite{chazy1911equations} and recently completed by C.Cosgrove \cite{cosgrove2000chazy} in case of a polynomial right-hand side. Finally for the order $n=4$ the polynomial problem was completed by C.Cosgrove \cite{cosgrove2000P2, cosgrove2006P1}. 

And although Painlev\'e classification has been successfully completed by the author for certain algebraic classes of equations of the arbitrary order, such as binomial-type equations \cite{sobolevsky2005binomial3, sobolevsky2006binomialN, sobolevsky2006mono} or arbitrary order differential equations with quadratic right-hand side \cite{sobolevsky2014arxiv}, the classification of ODE's of order $n\geq 5$ is not yet accomplished in any general enough case. 

In particular it was found that nonlinear equations
\begin{equation}
w^{(n)}=F(w,z),
\label{ODE2}
\end{equation}
where $F$ is algebraic in $w$ and locally analytic in $z$ never possess Painlev\'e property for $n\geq 3$ \cite{sobolevsky2001, sobolevskii2003algsing,sobolevsky2004stam,sobolevskii2005alg}.

Painlev\'e classification of the PDE's largely goes along the same lines as the classification of the ODE's. Second order semilinear PDE's
$$
A(z,t)w_{zz}+B(z,t)w_{zt}+C(z,t)w_{tt}=F(z,t,w,w_z,w_t)
$$
with the right-hand side $F$ rational in $w,w_z,w_t$ with the coefficients locally analytic in $z,t$ have been classified in \cite{hlavaty1992painleve, Cosgrove1993PDE1, Cosgrove1993PDE2}. Third order equations with a polynomial right-hand side considered by \cite{Martynov1994PDE3}.
Certain partial results were obtained for particular classes of equations of the fourth and some higher orders \cite{kulesh2006, misnik2013, maldonado2010integrability, kulesh2016, kulesh2017}.

So far there are no results on Painlev\'e classification known to the author that would concern broad classes of PDE's of an arbitrary order. While the methodology of Painlev\'e analysis for PDE's has its own specifics, in some cases it could be sufficient to simply consider their correspondent reductions to ODE's, enabling transfer of the corresponding classifications for the ODE's into the PDE domain. The present work is a first attempt of transferring recent results of the author on the arbitrary order ODE's to the PDE's.
 
\section{Quasilinear PDE depending only on the highest order derivatives}

Consider an autonomous quasilinear partial differential equation containing only the highest order derivatives 
\begin{equation}
\sum\limits_{\nu, \sum_i \nu_i=n}A_{\nu}(w)\frac{\partial^n w}{\partial x_1^{\nu_1}\partial x_2^{\nu_2}...\partial x_m^{\nu_m}}=P(w),
\label{eqMain}
\end{equation}
where $x=(x_1,x_2,...,x_m)$ is a vector of $m$ independent complex variables, $\nu=(\nu_1,\nu_2,...,\nu_m)$ are multi-indexes, and $P$ as well as $A_{\nu}$ are polynomials in $w$ without common roots $P(w^*)=A_{\nu}(w^*)=0, \forall \nu$ (i.e. equation is non-reducible), while either $deg P>1$ or $\exists \nu: deg A_{\nu}>0$ (i.e. equation is essentially nonlinear).

Example of such an equation of the order $n=3$ with $m=2$ independent variables $z,t$ might look like
$$
w_{ttt}+w_{zzz}=w^2.
$$
{\bf Theorem 1.} 
{\it Solutions of the equations of the class (\ref{eqMain}) always admit moveable critical singularities, so such equations do not possess Painlev\'e property.}\\
{\bf Proof.} One can see that all possible reductions of the equation (\ref{eqMain}) along the one-dimensional linear manifolds $z=c_1 x_1+c_2 x_2+...+c_m x_m$ could be written as
\begin{equation}
\left[\sum_\nu\left(\prod\limits_j c_j^{\nu_j}\right)A_{\nu}(w)\right]\frac{d^n w}{d z^n}=P(w),
\label{eqReducedC}
\end{equation}
i.e. takes the form
\begin{equation}
w^{(n)}=\frac{P(w)}{Q(w)},
\label{eqR}
\end{equation}
where $w=w(z)$ and its derivatives stand for ordinary derivatives in $z$, while $Q=\left[\sum_\nu\left(\prod\limits_j c_j^{\nu_j}\right)A_{\nu}(w)\right]$ is a polynomial in $w$. 

It can be seen that under certain choice of the constants $c_1,c_2,...,c_m$ the polynomial $Q$ could be guaranteed not to have common roots with $P$, while $deg Q=\max deg A_{\nu}$ (its major coefficient as well as values in all the roots of $P$ are linear combinations of non-zero major coefficients/values of $A_{\nu}$ and could be guaranteed non-zero for certain $c_1,c_2,...,c_m$). 

This way equation (\ref{eqR}) is essentially non-linear and belongs to the class (\ref{ODE2}). According to \cite{sobolevsky2001,sobolevsky2004stam} non-linear equations of the class (\ref{eqR}) admit moveable critical singularities. Then so does the corresponding solution 
$$
w=w(c_1 x_1+c_2 x_2+...+c_m x_m)
$$
of the original equation (\ref{eqMain}). The proof of the theorem is now complete.

\section{Conclusions}

For a broad class of autonomous quasilinear nonlinear partial differential equations of the arbitrary order $n\geq 3$, including only the highest order derivatives of order $n$, it has been proven that the solutions admit moveable critical singularities, i.e. no such equation could possess the Painlev\'e property. 

The present work provides an example and opens up potential for transferring recent results in the Painlev\'e classification of higher order ordinary differential equations  into the partial differential equation domain. And while direct reduction to ordinary differential equations used in this paper is not always sufficient for the Painlev\'e classification of partial differential equations, we believe this approach could largely help excluding the classes of equations that certainly do not possess Painlev\'e property, enabling other methods of Painlev\'e analysis for the remaining classes. Resulting Painlev\'e classifications could benefit multiple applied areas concerned with integrability of nonlinear evolutionary partial differential equations.



\end{document}